\documentclass[12pt]{article}
\begin{document}
\newcommand{\text}[1]{\mbox{#1}}
\newcommand{\Section}[1]{\section{#1}\setcounter{equation}{0}}
\renewcommand{\theequation}{\thesection.\arabic{equation}}

\title{On exact
solutions of the equation\\ for a pion in
external fields: III.}
\author{S.I.
Kruglov\thanks{E\_mail: skruglov23@hotmail.com}
\\ \it International Education
Centre, 2727 Steeles Ave. W, \# 202,
\\ \it Toronto, ON M3J 3G9, Canada
\\ \it
On leave from
\\ \it National Scientific Center of
Particle and High Energy
Physics,
\\ \it
M. Bogdanovich St. 153, Minsk 220010, Belarus}
\maketitle

\begin{abstract}
The exact solutions of the equation for a composite
scalar
particle which possesses electric and magnetic polarizabilities are
found in an
external uniform static magnetic field parallel to the direction
of propagation
of the quantized electromagnetic wave. As a particular case,
the relativistic
coherent states of a particle in the field of photons were
considered. When the
average photon number trends to infinity, the
solution tends to the solution
corresponding to  the
external classical electromagnetic wave with circular polarization.
\end{abstract}

\Section{Introduction}\label{sec:1}

In this paper we find and
discuss new solutions of the equation for a composite
scalar particle (pion or
kaon) with electromagnetic polarizabilities in
the case of the superposition of
the uniform static magnetic field and
quantized plane electromagnetic wave. In
\cite{1,2,3} exact solutions for a
particle which possesses
electromagnetic polarizabilities in the field of
the classical static uniform
electric, magnetic fields and the
electromagnetic wave were considered. The
advent of powerful lasers have been important to the understanding of
the interaction of charged particles with strong
light waves. In the neighborhood of neutron stars
and in the early stages of
the evolution of the universe, magnetic fields also
reach huge values. For
point-like particles such investigations were undertaken
in~\cite{4,5,6}. But it is
well known that scalar particles such as pions and
kaons have composite
nature and they possess electric $\alpha $ and magnetic
$\beta $
polarizabilities~\cite{7,8}. Therefore, it is interesting to learn the
effects
which are connected with the complex structure of particles. The case of
classical electromagnetic waves is justified when there are a lot of photons
in
light waves. This is a particular case of those where a particle interacts
with
definite numbers of separate photons. The latter is the most general case
and it can
be compared with the semiclassical approaches. The wavefunctions
of particles in
the field of a quantized electromagnetic wave are of some
interest when particles
interact with fields depending on time and the
number of absorbed or emitted
photons is indefinite.

In Section~\ref{sec:2} we find wavefunction of a pion
(or kaon) in a uniform
constant magnetic field and in the field of a quantized
monochromatic wave
with circular polarizations propagating along the magnetic
field. Section~\ref{sec:3}
contains the expansion of the found solutions in
small parameter $1/(k_0V)$ ($k_0$ is the energy of a photon and $V$ is 
the volume). In Section~\ref{sec:4} the
coherent wave with average number of photons
$<n>$ were constructed. The
limit when the photon number and the volume go to
infinity was investigated,
which corresponds to the case of the external
classical electromagnetic
wave. Section~\ref{sec:5} contains a conclusion.

\Section{A uniform magnetic field parallel to
a quantized electromagnetic wave}\label{sec:2}

The equation of motion for a composite scalar particle, taking into
account its electromagnetic polarizabilities, is given by~\cite{2, 3}
\begin{equation}
D_\mu ^2\varphi -D_\mu \left[ \left( D_\nu \varphi \right)
K_{\mu \nu
}\right] -m_{eff}^2\varphi =0
\label{(2.1)}\end{equation}
where the
squared effective mass is $m_{eff}^2=m^2\left( 1-\beta F_{\mu \nu
}^2/\left(
2m\right) \right) $, $m$ is the rest mass of a scalar particle,
$D_\mu =\partial
_\mu -ieA_\mu $ is the covariant derivative with the vector
potential $A_\mu $
and charge $e$, $K_{\mu \nu }=\left( \alpha +\beta
\right) F_{\mu \alpha }F_{\nu
\alpha }/m$ and $F_{\mu \nu }=\partial _\mu
A_\nu -\partial _\nu A_\mu $ is the
strength tensor. We use units in
which $\hbar =c=1$.

For the
superposition of the uniform static magnetic field and the quantized
monochromatic electromagnetic field we can choose the vector potential 
in the form
\begin{equation}
A_\mu =Hx_2e_{1\mu }+\frac 1{2\sqrt{k_0V}}\left[ \left(
e_{1\mu }-ie_{2\mu
}\right) c^{-}e^{i(kx)}+\left( e_{1\mu }+ie_{2\mu }\right)
c^{+}e^{-i(kx)}\right],
\label{(2.2)}\end{equation}
where $H$ is the strength of
a magnetic field, $e_{1\mu }$ and $e_{2\mu}$ are the unit vectors which
can be chosen as $e_{1\mu }=(1,0,0,0)$, 
$e_{2\mu }=(0,1,0,0)$, wavevector
$k_\mu =({\bf k},ik_0)$ so that ($e_1k)=\left( e_2k\right) =k^2=0$ and
$k_0$ is the photon energy, $\left(
kx\right) ={\bf k}{\bf x}-k_0x_0$ ($x_0$ is the time). The choice (2.2)
corresponds to the circular polarization of the
electromagnetic wave. The $V=L^3$ is the normalizing volume, where $L$ 
is the normalizing length so
that $k_0=2\pi /L$. For creation $c^{+}$ and annihilation
$c^{-}$ operators
we use the coordinate representation
\begin{equation}
c^{-}=\frac 1{\sqrt{2}}\left( \xi +\frac \partial {\partial \xi }\right),
\hspace{1.0in}
c^{+}=\frac 1{\sqrt{2}}\left( \xi -\frac \partial {\partial
\xi}\right)
\label{(2.3)}\end{equation}
with the commutation relation
$[c^{-},c^{+}]=1$. After the transformation of
the function $\varphi $~\cite{9}:
\begin{equation}
\varphi =U\chi \left( x_2,\xi \right) ,
\hspace{0.5in}
U=\exp \left\{
i\left[ \left( qx\right) +\frac{\left( kx\right) }2\left(
\partial _\xi^2-\xi ^2\right) \right] \right\} ,
\label{(2.4)}\end{equation}
where $\partial _\xi =\partial /\partial \xi $, $q_\mu =\left(
q_1,0,q_3,iq_0\right) $ is a quasi-momentum of a particle, and
equation~(\ref{(2.1)}) with
taking into account~(\ref{(2.2)}) is transformed to
\begin{eqnarray}
\left[ l\left( \partial _\xi ^2-\xi ^2\right) +q_1d_1\xi
-2hbAx_2\xi
+2hq_1Ax_2-h^2Ax_2^2+C\right] \chi&&\\
 -d_1\frac{\partial ^2\chi
}{\partial \xi
\partial x_2}+A\frac{\partial ^2\chi }{(\partial
x_2)^2}&=&0 ,\nonumber
\label{(2.5)}\end{eqnarray}
where we have introduce the notations
\begin{equation}
\begin{array}{lll}
l=b^2A-\left( qk\right) B , &
A=1-WH^2 , & W=\frac{\alpha +\beta }m , \\
B=1+\frac{W\left( qk\right) }{2k_0V} ,
&  d_1=2bA-H\left(qk\right) f , &  f=W\sqrt{\frac 2{k_0V}} , \\
h=eH ,&
C=q_0^2-q_1^2A-q_3^2-m_{eff}^2 ,& b=\frac{e}{\sqrt{2k_0V}} .
\end{array}
\label{(2.6)}\end{equation}

We have the photon $\xi $ and coordinate $x_2$
variables which enter
equation~(\ref{(2.5)}). To separate variables in
equation~(\ref{(2.5)})
we need to make linear transformations of $\xi $, $x_2$
and to introduce
new variables
\begin{equation}
\eta _1=a_2\left( \xi
+a_1x_2\right) , 
\hspace{1.0in} 
\eta _2=a_3\left(x_2+\frac{d_1-2Aa_1}{2l-d_1a_1}\xi \right) . 
\label{(2.7)}\end{equation}

After transformation~(\ref{(2.7)}) the operator
\begin{eqnarray}
l\partial _\xi
^2-d_1\frac{\partial ^2}{\partial \xi \partial x_2}+A\frac{
\partial
^2}{(\partial x_2)^2}&&\\
&&\hspace{-1.6in}=\left( l+Aa_1^2-d_1a_1\right) \left[
a_2^2
\frac{\partial ^2}{\left( \partial \eta _1\right)
^2}+a_3^2\frac{4lA-d_1^2}{
\left( 2l-d_1a_1\right) ^2}\frac{\partial ^2}{\left(
\partial \eta _2\right)
^2}\right]\nonumber
\label{(2.8)}\end{eqnarray}
is
diagonal and $a_1,$ $a_2$, $a_{3}$ are free parameters. The
requirement that the quadratic form
\[
l\xi ^2-q_1d_1\xi +2hbAx_2\xi -2hq_1Ax_2+h^2Ax_2^2,
\]
upon using equation~(\ref{(2.7)}), does not contain the term
$\eta _1\eta
_2,$ gives the following restriction on the coefficient $a_1$:
\begin{equation}
a_1^2\left( 2bhA^2-ld_1\right) +2\left( l^2-A^2h^2\right) a_1+hA\left(
hd_1-2bl\right) =0 .
\label{(2.9)}\end{equation}

The transformations~(\ref{(2.7)}) with the solutions of equation~(\ref{(2.9)})
guarantee that~(\ref{(2.5)}) becomes diagonal and variables
$\eta _1$, $\eta _2$
will be separated. So we have two free parameters
$a_2$, $a_3$. This freedom
will be used to simplify the solutions of
equation~(\ref{(2.5)}).
Using~(\ref{(2.7)}) and~(\ref{(2.9)}),
equation~(\ref{(2.5)}) is converted to
\begin{eqnarray}
&&
\hspace{-2em}
\left[\frac{M-a_1N}{2}\left( a_2^2\frac{\partial^2}{\left(
  \partial \eta _1\right)^2}+a_3^2\frac{4lA-d_1^2}{M^2}\frac{\partial ^2}
  {\left( \partial \eta_2\right) ^2}\right) 
\right.\nonumber
\\&&
-\frac{lM-bhAN}{a_2^2\left(M-a_1N\right) }\eta _1^2
-\frac{M^2\left(
la_1^2+h^2A-2bhAa_1\right) }{a_3^2\left( M-a_1N\right) ^2}\eta _2^2
\label{(2.10)}\\&&
+\left.\frac{q_1\left( Md_1-2AhN\right) }{a_2\left( M-a_1N\right) }\eta
_1+\frac{
q_1M\left( 2Ah-d_1a_1\right) }{a_3\left( M-a_1N\right) }\eta _2+
C\right]\chi \left( \eta _1,\eta _2\right) =0 ,
\nonumber\end{eqnarray}
where $M=2l-a_1d_{1}$, $N=d_1-2Aa_1.$ We
can look for the solution
to equation~(\ref{(2.10)}) in the form
\begin{equation}
\chi \left( \eta _1,\eta _2\right) =\chi _1\left( \eta
_1\right) \chi
_2\left( \eta _2\right) ,
\label{(2.11)}\end{equation}
where wavefunctions $\chi _1\left( \eta _1\right) $, $\chi _2\left( \eta
_2\right) $ are
the solutions of the equations
\begin{eqnarray}
&&
\left[
a_2^2\frac{M-a_1N}2\frac{\partial ^2}{\left( \partial \eta _1\right)^2}
\right.\label{(2.12)} \\&& \left.\hspace{2em}
-\frac{lM-bhAN}{a_2^2\left(
M-a_1N\right) }\eta _1^2+\frac{q_1\left(
Md_1-2AhN\right) }{a_2\left(
M-a_1N\right) }\eta _1\right] \chi _1=\lambda
_1\chi _1\nonumber
\\ &&
\left[
a_3^2\frac{M-a_1N}2\frac{4lA-d_1^2}{.M^2}\frac{\partial ^2}{\left(
\partial \eta
_2\right) ^2}\right.\label{(2.13)} \\&& \left.\hspace{2em}
-\frac{M^2\left(
la_1^2+h^2A-2bhAa_1\right) }{
a_3^2\left( M-a_1N\right) ^2}\eta
_2^2+\frac{q_1M\left( 2Ah-d_1a_1\right) }{
a_3\left( M-a_1N\right) }\eta
_2\right] \chi _2=\lambda _2\chi _2\nonumber
\end{eqnarray}
and the eigenvalues
$\lambda _{1}$, $\lambda _2$ are related by
\begin{equation}
\lambda
_1+\lambda _2=-C=q_1^2A+q_3^2+m_{eff}^2-q_0^2 .
\label{(2.14)}\end{equation}
Equations~(\ref{(2.12)}),~(\ref{(2.13)}) can be simplified by a shift in the
variables 
$\eta_1$, $\eta_2$:
\begin{equation}
\overline{\eta }_1=\eta
_1-\frac{q_1\left( Md_1-2AhN\right) }{a_2^3\left(
M-a_1N\right) ^2} ,
\hspace{0.25in} \overline{\eta }_2=\eta _2-\frac{
q_1M^3\left( 2Ah-d_1a_1\right)
}{a_3^3\left( M-a_1N\right) ^2\left(
4lA-d_1^2\right) }
\label{(2.15)}\end{equation}
with the parameters
\begin{equation}
a_2=\frac{\left[ 2\left( lM-bhAN\right) \right] ^{1/4}}{\sqrt{M-a_1N}} ,
\hspace{0.25in} 
a_3=\frac{M\left[ 2\left( la_1^2+h^2A-2bhAa_1\right) \right]
^{1/4}}{\left( M-a_1N\right) ^{3/4}\left( 4lA-d_1^2\right) ^{1/4}} .
\label{(2.16)}\end{equation}

Taking into
account~(\ref{(2.15)}),~(\ref{(2.16)}), equations~(\ref{(2.12)}),~(\ref{(2.13)})
become
\begin{eqnarray}
\left( \frac{\partial ^2}{\partial \overline{\eta
}_1^2}-\overline{\eta }
_1^2+\varepsilon _1\right) \chi _1\left( \overline{\eta
}_1\right) &=&0 ,
\label{(2.17)}\\
\left( \frac{\partial ^2}{\partial
\overline{\eta }_2^2}-\overline{\eta }
_2^2+\varepsilon _2\right) \chi _2\left(
\overline{\eta }_2\right) &=&0 ,
\label{(2.18)}\end{eqnarray}
where
\begin{eqnarray}
\varepsilon _1&=&\frac{q_1^2\left( d_1M-2AhN\right) ^2}{\left[
2\left(
lM-bhAN\right) \right] ^{3/2}\left( M-a_1N\right) }-\frac{2\lambda
_1}{\sqrt{
2\left( lM-bhAN\right) }} ,\label{(2.19)}\\
\varepsilon
_2&=&\frac{q_1^2\left( 2Ah-d_1a_1\right) ^2\sqrt{\left(
M-a_1N\right) }}{\left[
2\left( la_1^2+h^2A-2bhAa_1\right) \right] ^{3/2}
\sqrt{4lA-d_1^2}}-\frac{2\lambda _2\sqrt{\left( M-a_1N\right) }}{\sqrt{
2\left(
la_1^2+h^2A-2bhAa_1\right) \left( 4lA-d_1^2\right) }}.\nonumber
\end{eqnarray}
The solutions of equations~(\ref{(2.17)}),~(\ref{(2.18)}) are well
known~\cite{10}.
The finite solutions of equations~(\ref{(2.17)}),~(\ref{(2.18)})
at $\overline{\eta }_1$,
$\overline{\eta}_2\rightarrow \infty $ are
\begin{equation}
\chi _1\left( \overline{\eta
}_1\right) =N_1H_n\left( \overline{\eta }
_1\right) \exp \left( -\frac
12\overline{\eta }_1^2\right) ,
\hspace{0.1in}
\chi _2\left( \overline{\eta
}_2\right) =N_2H_s\left( \overline{\eta }
_2\right) \exp \left( -\frac
12\overline{\eta }_2^2\right) ,
\label{(2.20)}\end{equation}
where $N_1,N_2$ are
normalization constants, $H_n\left( \overline{\eta }
_1\right) $, $H_s\left(
\overline{\eta }_2\right) $ are the Hermite
polynomials and values $\varepsilon
_1$, $\varepsilon _2$ are connected with
quantum numbers $n$, $s$:
\begin{equation}
\varepsilon _1=2n+1, \hspace{0.3in}
\varepsilon _2=2s+1,
\hspace{0.3in}
(n=1,2,3..., s=1,2,3....)
\label{(2.21)}\end{equation}

From
equations~(\ref{(2.11)}),~(\ref{(2.20)}) we can write out the solution of 
equation~(\ref{(2.10)})
\begin{equation}
\chi \left( \overline{\eta
}_1,\overline{\eta }_2\right) =N_0H_n\left( 
\overline{\eta }_1\right) H_s\left(
\overline{\eta }_2\right) \exp \left[
-\frac 12\left( \overline{\eta
}_1^2+\overline{\eta }_2^2\right) \right],
\label{(2.22)}\end{equation}
and the
constant $N_0$ found from the normalization condition is
\begin{equation}
N_0=\left( \pi 2^{n+s}n!s!\right) ^{-1/2} .
\label{(2.23)}\end{equation}

The new
variables $\overline{\eta }_1$, $\overline{\eta }_2$,
with the help
of~(\ref{(2.7)}),~(\ref{(2.15)},~(\ref{(2.16)}), take the form
\begin{eqnarray}
\overline{\eta }_1&=&\frac{\left[ 2\left( lM-bhAN\right) \right] ^{1/4}}{
\sqrt{M-a_1N}}\left( \xi +a_1x_2\right)
\nonumber\\&&\hspace{.2in}
-\frac{q_1\left( d_1M-2AhN\right) }{
\left[ 2\left( lM-bhAN\right) \right]
^{3/4}\sqrt{M-a_1N}} ,
\label{(2.24)}\\
\overline{\eta }_2&=&\frac{M\left[ 2\left(
la_1^2+h^2A-2bhAa_1\right) \right]
^{1/4}}{\left( M-a_1N\right) ^{3/4}\left(
4lA-d_1^2\right) ^{1/4}}\left(
x_2+\frac NM\xi \right) 
\nonumber\\&&\hspace{.2in}
-\frac{q_1\left( 2Ah-d_1a_1\right) \left(
M-a_1N\right) ^{1/4}}{\left[ 2\left( la_1^2+h^2A-2bhAa_1\right) \right]
^{3/4}\left( 4lA-d_1^2\right) ^{1/4}} .
\label{(2.25)}\end{eqnarray}

In equation~(\ref{(2.22)}) the photon variable $\xi $ is mixed with the
'magnetic' variable $x_2$. We will see that quantum numbers $n$ and $s$
describe the photon number and the principal quantum number.
From~(\ref{(2.19)}),~(\ref{(2.21)}),~(\ref{(2.14)}) we find the dispersion
relation for the
quasimomentum $q_\mu $ of a composite scalar particle
interacting with the
quantized monochromatic electromagnetic wave:
\begin{eqnarray}
q_0^2&=&q_1^2\left\{ A-\frac 14\left[ \frac{\left(
d_1M-2AhN\right) ^2}{
\left( lM-bhAN\right) \left( M-a_1N\right) }+\frac{\left(
2Ah-d_1a_1\right)
^2}{\left( la_1^2+h^2A-2bhAa_1\right) }\right] \right\}
\nonumber\\&&
+q_3^2+m_{eff}^2
-\left( 2n+1\right) \sqrt{\frac{lM-bhAN}2}
\label{(2.26)}\\&&
-\left( 2s+1\right) \sqrt{\frac{
\left(
la_1^2+h^2A-2bhAa_1\right) \left( 4lA-d_1^2\right) }{2\left(
M-a_1N\right) }} .
\nonumber\end{eqnarray}

It should be noted that the squared root entering
equation~(\ref{(2.26)}) has two
signs $\pm $ and equation~(\ref{(2.26)}) is one
of the possible solutions for the
quasimomentum of a scalar particle.
Equation~(\ref{(2.26)}) for the full momentum of the composite scalar particle
interacting with photons is very complicated as the four-vector $q_\mu $ enters
the quantities $l$, $B$, $d_1$, $M$, $N$ (see~(\ref{(2.6)})). To clear up the
physical
meaning of quantum numbers $n$, $s$, let us consider the particular
case
when coupling constant $e=0$, i.e. no interaction between a particle and
the
quantized electromagnetic wave. In addition we consider point-like
particles, when $\alpha $, $\beta \rightarrow 0$ and a consequence $
W\rightarrow 0$, $A,B\rightarrow 1$, $a_1,N,d_1\rightarrow 0$, $M\rightarrow
2l$. Then equation~(\ref{(2.26)}) is simplified to
\begin{equation}
q_0^2=q_3^2+m^2-\left( 2s+1\right) eH-\left( 2n+1\right) \left( qk\right) .
\label{(2.27)}\end{equation}

If the photons move along the $x_3$-direction so
that $k_1=k_2=0$, $k_3=k_0$,
the solution of equation~(\ref{(2.27)}) is
\begin{equation}
q_0=\left( n+\frac 12\right) k_0+\sqrt{\left[ q_3-\left(
n+\frac 12\right)
k_0\right] ^2+m^2-\left( 2s+1\right) eH} .
\label{(2.28)}\end{equation}

The first term $\left( n+1/2\right) k_0$ in
equation~(\ref{(2.28)}) is the energy of 
$n$ photons. Now we introduce the
third component of a particle momentum $p_3=q_3-\left( n+1/2\right) k_0$ 
and take into account the Landau energy of
a scalar particle in the external
magnetic field
\begin{equation}
p_0^2=p_3^2+m^2-\left( 2s+1\right) eH .
\label{(2.29)}\end{equation}

Then equation~(\ref{(2.28)}) represents the energy
of a quasiparticle as a sum of an
energy of $n$-photons and the energy of a
particle in the uniform static
magnetic field. Therefore $s$ is the principal
quantum number and $n$ is the
number of photons in the quantized monochromatic
electromagnetic wave. So in
general case the four-vector $q_\mu$is the
quasimomentum of a
composite scalar particle in the external magnetic field
parallel to a
direction of the propagation of a quantized electromagnetic wave.
\Section{Approximate solutions at $V\rightarrow\infty$}\label{sec:3}

The interesting physical case is when the volume of the system goes to
infinity. We will expand all quantities in the power of $1/V$ and leave
only terms in order
of $1/V$, $W/V$ and we will neglect the small terms $1/V^2$, $W^2$ and 
so on. Terms which contain $W$ describe the complex structure of a
scalar particle
which possesses electromagnetic polarizabilities $\alpha $
and $\beta $. In this
approximation with the accuracy of $O\left( 1/V\right)$, 
$O\left( W/V\right) $ we have the values of parameters~(\ref{(2.6)})
\begin{eqnarray}
l&=&-Q+\frac
1{2k_0V}\left( e^2A-WQ^2\right) ,
\nonumber\\
d_1&=&\sqrt{\frac 2{k_0V}}\left(
eA-WHQ\right) ,
\label{(3.1)}\\
M&=&-2Q-\frac 1{k_0V\left( h-Q\right) }
  \left[
e^2Q-\frac{W\left(2h^4Q-3h^3Q^2-h^2Q^3+hQ^4-Q^5\right) }{h^2-Q^2}\right] ,
\nonumber\\
N&=&-\sqrt{\frac 2{k_0V}}\frac 1{h-Q}\left[
eQ+\frac{WH\left(h^2Q^2+Q^4\right) }{h^2-Q^2}\right] ,
\nonumber\end{eqnarray}
and the solution of equation~(\ref{(2.9)}) is
\begin{equation}
a_1=\frac{eh}{\sqrt{2k_0V}\left( h-Q\right) }\left( 1+\frac{2WH^2Q^2}{
h^2-Q^2}\right) ,
\label{(3.2)}\end{equation}
where $Q=\left( qk\right) $,
$h=eH$. Inserting~(\ref{(3.1)}),~(\ref{(3.2)})
into~(\ref{(2.26)}) we
find in
the considered approximation the dispersion relation
\begin{equation}
q_0^2=q_3^2+m_{eff}^2-\left( 2n+1\right) Q\left( 1+\delta _1\right) -\left(
2s+1\right) h\left( A+\delta _2\right)
\label{(3.3)}\end{equation}
with the
values $\delta _{1\text{,}}$ $\delta _2$,
\begin{eqnarray}
\delta _1&=&\frac
1{2k_0V\left( h-Q\right) }\left[ e^2-\frac{W\left(
h^3+2hQ^2-Q^3-3h^2Q\right)
}{h-Q}\right] ,
\label{(3.4)}\\
\delta _2&=&-\frac 1{2k_0V\left( h-Q\right)
}\left[ e^2-\frac{Wh\left(
h^3+3hQ^2+2Q^3-4h^2Q\right) }{2Q\left( h-Q\right)
}\right] .
\label{(3.5)}\end{eqnarray}

It should be noticed that the term with
$q_1$ in~(\ref{(3.3)}) is absent and it is
possible to find a solution to it. As
the values of $\delta_1$ and $\delta_2$
are small at $V\rightarrow \infty $, we
can introduce the momentum $p_\mu$ of a scalar
composite particle in an external
magnetic field,
\begin{equation}
p^2=-m_{eff}^2+\left( 2s+1\right) h\left(
A+\delta _2\right) , \hspace{0.5in}
(p^2=p_3^2-p_0^2) .
\label{(3.6)}\end{equation}

Expression~(\ref{(3.6)}) at $\delta _2\rightarrow
0$ is transformed into the known
relation~\cite{2} for the momentum of a
composite particle in the pure uniform
magnetic field. Then
equation~(\ref{(3.3)}) has the solution
\begin{equation}
q_\mu =p_\mu +\left(
n+\frac 12\right) k_\mu \left( 1+\delta _1\right) ,
\hspace{0.5in} \left( \mu
=0,3\right) .
\label{(3.7)}\end{equation}

We took into account that $Q=\left(
qk\right) =\left( pk\right) $ as $k_\mu
^2=0$. So the quasienergy of a particle
$q_0$ depends on the number of photons $
n $, but $q_1$ does not.
Equation~(\ref{(3.7)}) determines an energy $q_0$
of a system which consists of
a scalar particle and $n$ photons. This energy
depends on the quantum number
$p_3$. We notice that physical meaning of a
variable $p_0$ as a particle energy
in the external magnetic field occurs in
accordance with~(\ref{(3.6)}) only at
$V\rightarrow \infty $ when
$\delta_2\rightarrow 0$ because $\delta _2$ depends
on $Q=\left( pk\right) $
(see~(\ref{(3.5)})).
Expressions~(\ref{(3.1)},~(\ref{(3.2)}) have a singularity at $h=\pm Q$ or
$eH=\pm
\left( pk\right) $. This corresponds to the case of resonance when the
photon energy $k_0$ equals to the cyclotron frequency $\Omega =eH/\left(
p_0-p_3\right) $. So all values become incorrect near the resonance $
k_0=\Omega $ (see also~\cite{5}). Here we consider single-particle theory (the
scalar field is not a quantized field) and therefore there are some
difficulties
to construct a wavefunction in the vicinity of a resonance.

The variables
$\eta _1$, $\eta _2$ in~(\ref{(2.24)}),~(\ref{(2.25)}), in the present
approximation, transform into
\begin{eqnarray}
\overline{\eta }_1&=&\left[
1+\frac{e^2h}{4k_0V\left( h-Q\right) ^2}+W\frac{
h^5-h^4Q+3h^3Q^2+h^2Q^3}{4k_0V\left( h-Q\right) ^2\left( h^2-Q^2\right) }
\right] \xi
\nonumber\\&&
+\frac{eh}{\sqrt{2k_0V}\left( h-Q\right) }\left[
1+\frac{2WH^2Q^2
}{h^2-Q^2}\right] x_2
\label{(3.8)}\\&&
-\frac{q_1}{\sqrt{2k_0V}\left( h-Q\right) }\left[ e+WH\frac{
hQ^2+h^3-Qh^2+Q^3}{h^2-Q^2}\right]
\nonumber\end{eqnarray}
and
\begin{eqnarray}
\overline{\eta }_2&=&\sqrt{h}\left[ 1+\frac{e^2h}{4k_0V\left( h-Q\right) ^2}
+Wh\frac{h^5-2h^4Q+2h^3Q^2+4h^2Q^3+5Q^4h-2Q^5}{8k_0VQ\left( h-Q\right)
^2\left(
h^2-Q^2\right) }\right] x_2
\nonumber\\&&
+\frac{\sqrt{h}}{\sqrt{2k_0V}\left(
h-Q\right) }\left[ e+WHQ\frac{h^2+Q^2}{
h^2-Q^2}\right] \xi
\label{(3.9)}\\&&
-q_1\left[ \frac 1{\sqrt{h}}+\frac{\sqrt{h}e^2}{
4k_0V\left( h-Q\right)
^2}+\frac{W\left(
h^6-2h^5Q+6h^4Q^2+h^2Q^4+2hQ^5\right) }{8\sqrt{h}Qk_0V\left(
h-Q\right)
^2\left( h^2-Q^2\right) }\right].\nonumber
\end{eqnarray}

It can be
seen from~(\ref{(3.9)}) that there are singularities at $h\rightarrow 0$, $
Q\rightarrow 0$. Therefore, we can not consider here the cases of a pure
magnetic
field and quantized electromagnetic wave. Solutions for pions in
external
electromagnetic fields of such configurations were learned in
details~\cite{1,2}.

With the help
of~(\ref{(2.4)}),~(\ref{(2.22)}),~(\ref{(2.23)} we write out the final solution
to
equation~(\ref{(2.1)}) for pions (or other composite scalar particles) in the
external uniform static magnetic field parallel to quantized electromagnetic
wave
\begin{eqnarray}
\varphi \left( x,\xi \right) &=&\left( \pi
2^{n+s}n!s!\right) ^{-1/2}
\label{(3.10)}\\&&
\hspace{-1em}\times\exp
\left\{
i\left[ \left( qk\right) +\frac 12\left( kx\right) \left( \partial
_\xi ^2-\xi
^2\right) \right] -\frac 12\left( \overline{\eta }_1^2+\overline{
\eta
}_2^2\right) \right\} H_n\left( \overline{\eta }_1\right) H_s\left( 
\overline{\eta }_2\right) .
\nonumber\end{eqnarray}

The exponential operator
in~(\ref{(3.10)}) acts on the Hermite polynomials and on
the exponent which
contains variables $\overline{\eta }_1$, $\overline{\eta }_2$.
Terms with the
variable $W$ in~(\ref{(3.9)}) describe the contribution from the
internal
structure of a composite scalar particle. If we put $W=0$ in~(\ref{(3.9)})
we
will have the wavefunction~(\ref{(3.10)}) of a point-like scalar particle in
the
external magnetic field parallel to a quantized electromagnetic wave. But
real scalar particles (pions or kaons) have the complex quark structure and
therefore we should take into account all terms in~(\ref{(3.8)}),~(\ref{(3.9)}). 
\Section{Coherent states of electromagnetic waves}\label{sec:4}

The exact
solution~(\ref{(3.10)}) gives the wavefunction of the charged scalar
composite
particle with the arbitrary phase of the wave interacting with $n$
separate
external photons. The linear combination of solutions~(\ref{(3.10)}) with
different photon numbers $n$ and phases is also the solution of
equation~(\ref{(2.1)}).
Here we consider the Poisson distribution of the numbers
$n$ which
corresponds to the coherent external electromagnetic wave.

The wave
packet for such states is described by the sum~\cite{11}
\begin{equation}
\varphi _{<n>}(x,\xi )=\exp \left( -\frac 12<n>\right) \sum_{n=0}^\infty
\frac
1{\sqrt{n!}}<n>^{n/2}\varphi (x,\xi ) , 
\label{(4.1)}\end{equation}
where $<n>$
is the average photon number. Here we have chosen the phase of
the
electromagnetic wave which equals zero.

Inserting~(\ref{(3.10)}) into
(\ref{(4.1)}) and using the properties of Hermite
polynomials~\cite{10}
\begin{equation}
\sum_{m=0}^\infty \frac{t^m}{m!}H_m(x)=\exp \left(
-t^2+2tx\right)
\label{(4.2)}\end{equation}
we find
\begin{equation}
\varphi
_{<n>}(x,\xi )=\left( \pi 2^ss!\right) ^{-1/2}\sqrt{z_1}\exp \left[
-\frac
12<n>(1+z_1^2)+i\left( px\right) +i\left( q_1x_1\right) \right] G_s ,
\label{(4.3)}\end{equation}
where $z_1=exp[i(kx)(1+\delta _1)]$,
$(px)=p_3x_3-p_ox_o$ and we introduced
the notation
\begin{equation}
G_s=U_1H_s(\eta _2)\exp \left[ z_1\sqrt{2<n>}\overline{\eta }_1-\frac
12\left(
\overline{\eta }_1^2+\overline{\eta }_2^2\right) \right] ,
\label{(4.4)}\end{equation}
where $U_1=\exp \left[ i\frac 12\left( kx\right)
\left( \partial _\xi ^2-\xi
^2\right) \right] .$ The relation~(\ref{(3.7)}) was
used here for the four-vector
$q_\mu =\left( q_1,0,q_3,iq_0\right) $. To calculate
the action of the
operator $U_1$ on the Hermite polynomial and the exponential
function in~(\ref{(4.4)})
we use the expansion
\begin{equation}
H_s(\overline{\eta }_2)\exp \left[ z_1\sqrt{2<n>}\overline{\eta }_1-\frac
12\left( \overline{\eta }_1^2+\overline{\eta }_2^2\right) \right]
=\sum_{m=0}^\infty \beta _m^sH_m(\xi )\exp \left( -\frac 12\xi ^2\right),
\label{(4.5)}\end{equation}
where
\begin{equation}
\beta _m^s=\frac
1{2^m\sqrt{\pi }m!}\int_{-\infty }^\infty d\overline{\xi }
H_s(\overline{\eta
}_2)H_m(\overline{\xi })\exp \left[ z_1\sqrt{2<n>}
\overline{\eta }_1-\frac
12\left( \overline{\eta }_1^2+\overline{\eta }_2^2+
\overline{\xi }^2\right)
\right].
\label{(4.6)}\end{equation}

With the help of~(\ref{(4.5)}) and the
property of the oscillator wavefunction
\begin{equation}
(\xi ^2-\partial _\xi
^2)H_m(\xi )\exp \left( -\frac 12\xi ^2\right)
=(2m+1)H_m(\xi )\exp \left(
-\frac 12\xi ^2\right)
\label{(4.7)}\end{equation}
equation~(\ref{(4.4)})
transforms into the expression
\begin{equation}
G_s=\sum_{m=0}^\infty \exp
\left[ -i\left( m+\frac 12\right) (kx)\right]
\beta _m^sH_m(\xi )\exp \left(
-\frac 12\xi ^2\right).
\label{(4.8)}\end{equation}

This function includes
oscillator wave functions with the different numbers
of $m$. It is possible to
calculate the sum by inserting coefficient~(\ref{(4.6)})
into~(\ref{(4.8)}) and
taking into account the relationship~\cite{10}
\begin{equation}
\sum_{m=0}^\infty \frac{z^m}{2^mm!}H_m(x) H_m(y)=\frac 1{\sqrt{1-z^2}
}\exp
\left[ \frac{2xyz-(x^2+y^2)z^2}{1-z^2}\right].
\label{(4.9)}\end{equation}

As a
result the function~(\ref{(4.8)}) becomes
\begin{equation}
G_s=\sqrt{\frac z{\pi
\left( z^2-1\right) }}\exp \left[ -\frac{\xi ^2(z^2+1)
}{2(z^2-1)}\right] I,
\label{(4.10)}\end{equation}
where the integral $I$ is given by
\begin{equation}
I=\int_{-\infty }^\infty d\overline{\xi }H_s(\overline{\eta }_2)\exp \left[
z_1\sqrt{2<n>}\overline{\eta }_1-\frac{\overline{\eta} _1^2+
\overline{\eta} _2^2}2-\frac{\overline{
\xi }^2(z^2+1)-4\xi \overline{\xi }z}{2(z^2-1)}\right],
\label{(4.11)}\end{equation}
and $z=exp\left[ i\left( kx\right) \right] .$ It
should be noted that
variables $\overline{\eta }_1$, $\overline{\eta }_2$ depend
on $\overline{
\xi }$ by the relationships~(\ref{(3.8)}),~(\ref{(3.9)}), where
instead of $\xi $ we have to
use $\overline{\xi }$. From
equations~(\ref{(3.8)}),~(\ref{(3.9)}) we can define the
relations
\begin{equation}
\overline{\eta }_1=R\overline{\eta }_2+P , \hspace{1.0in}
\overline{\xi }=b
\overline{\eta }_2+c,
\label{(4.12)}\end{equation}
where the
values $R$, $P$, $b$, $c$ can be easily found from~(\ref{(3.8)}),~(\ref{(3.9)}).
Using the relationships~(\ref{(4.2)}) the integral~(\ref{(4.11)}) is rewritten
as fallows:
\begin{eqnarray}
I&=&b\exp \left[
z_1\sqrt{2<n>}P-\frac{P^2}2+\frac{2\xi zc}{z^2-1}-\frac{
c^2(z^2+1)}{2(z^2-1)}\right] I_1,
\label{(4.13)}\\
I_1&=&\int_{-\infty }^\infty
d\overline{\eta }_2H_s(\overline{\eta }_2)\exp
\left( -X\overline{\eta
}_2^2+Y\overline{\eta }_2\right),
\label{(4.14)}\end{eqnarray}
where
\begin{eqnarray}
X&=&\frac 12\left[ 1+R^2+\frac{b^2(z^2+1)}{(z^2-1)}\right],
\label{(4.15)}\\
Y&=&\sqrt{2<n>}z_1R-RP-\frac{bc(z^2+1)}{z^2-1}-\frac{2\xi
zb}{z^2-1}.
\nonumber\end{eqnarray}
Taking into account~(\ref{(4.2)}), one
arrives at the equation
\begin{eqnarray}
\int_{-\infty }^\infty d\overline{\eta
}_2&&\exp \left( -t^2+2t\overline{\eta 
}_2-X\overline{\eta
}_2^2+Y\overline{\eta }_2\right) 
\nonumber\\&&
=\sum_{s=0}^\infty 
\frac{t^s}{s!}\int_{-\infty }^\infty d\overline{\eta }_2H_s(\overline{\eta }
_2)\exp \left( -X\overline{\eta }_2^2+Y\overline{\eta }_2\right) . 
\label{(4.16)}\end{eqnarray}

Calculating the integral in the left-hand site of
equation~(\ref{(4.16)}) and comparing
it with the right-hand site of (4.16) we
find the integral $I_1$:
\begin{equation}
I_1=\sqrt{\frac \pi X}\left(
\frac{X-1}X\right) ^{s/2}H_s\left( \frac Y{2
\sqrt{X(X-1)}}\right) \exp \left(
\frac{Y^2}{4X}\right).
\label{(4.17)}\end{equation}

The leading terms of
variables $R$, $P$, $b$, $c$ which enter~(\ref{(4.15)}) and
solution~(\ref{(4.17)}) are found from equations~(\ref{(3.9)}),~(\ref{(4.12)}):
\begin{eqnarray}
R&=&b+\frac{\sqrt{h}e}{2\sqrt{2k_0V}(h-Q)}\left(
1+WH^2\frac{h^2-2hQ+3Q^2}{h^2-Q^2}\right) ,
\nonumber\\
b&=&\frac{\sqrt{2k_0V}\left( h-Q\right) }{e\sqrt{h}}\left(
1-\frac{WHQ}e\frac{h^2+Q^2}{h^2-Q^2}\right) ,
\nonumber\\
P&=&c+\frac
e{2\sqrt{2k_0V}(h-Q)}
\label{(4.18)}\\&&\hspace{-2em}\times
\left\{
hx_2-q_1+\frac{WH}{e\left( h^2-Q^2\right) }\left[ x_2h^2\left(Q^2+2hQ-h^2\right)
-q_1\left( h^3-Q^2h+2Q^3\right) \right] \right\} ,
\nonumber\\
c&=&b\left(
\frac{q_1}{\sqrt{h}}-\sqrt{h}x_2\right).
\nonumber\end{eqnarray}

Taking into
account equation~(\ref{(4.18)}), the quantities $X,Y$ in~(\ref{(4.15)})
entering~(\ref{(4.17)}) become
\begin{eqnarray}
X&=&\frac{b^2z^2}{z^2-1},
\label{(4.19)}\\
Y&=&b\left[ \sqrt{2<n>}z-\frac{2cz^2}{z^2-1}+\frac{2\xi
z}{z^2-1}+\frac{
WH\left( h^2x_2+q_1Q\right)
}{\sqrt{2k_0V}(h+Q)}\right]\nonumber
\end{eqnarray}
in the $V\rightarrow \infty
$ limit.

From~(\ref{(4.10)}),~(\ref{(4.13)}),~\ref{(4.17)}) and~(\ref{(4.19)})
we
write out the solution~(\ref{(4.3)}) when the number of photons $n$ and
the
volume of the quantization of electromagnetic waves $V$ go to
infinity (but
$\frac{<n>}{V}=const)$:
\begin{eqnarray}
\lim_{<n>\rightarrow \infty }\varphi
_{<n>}&=&
\nonumber\\&&\hspace{-1in}
\left( \pi 2^ss!\right)^{-1/2}H_s\left[
\sqrt{h}x_2-\frac{q_1}{\sqrt{h}}+
i\frac{a_1\sqrt{h}}{\sqrt{2}(h-Q)}(e+WHQ\frac{h^2+Q^2}{h^2-Q^2})\sin (kx)\right]
\nonumber\\&&\hspace{-1in}
 \times \exp \left\{\rule{0mm}{4ex}\right.  
i(px+q_1x_1)-\frac{\left( \xi -\sqrt{2\left\langle n\right\rangle }\right) ^2}2   
+iWH\frac{a_1}{\sqrt{2}}\frac{h^2x_2+q_1Q}{h+q}\sin (kx)
\nonumber\\&&\hspace{-.4in}
 -WH\frac{h-Q}{e(h+Q)}\left( \frac{q_1}h-x_2\right)
\left(h^2x_2+q_1Q\right)
\nonumber\\&&\hspace{-.4in}
 +WH\frac \xi
{\sqrt{2k_0V}}\frac{h^2x_2+q_1Q}{h+Q}\exp\left[ -i(kx)\right]
\left.\rule{0mm}{4ex}\right\},
\label{(4.20)}\end{eqnarray}
where $h=eH$,
$Q=(pk)$. The last term in~(\ref{(4.20)}) remains finite as the photon
variable
$\xi $ goes to $\sqrt{2<n>}$ at $<n>\rightarrow \infty $. Wavefunction
~(\ref{(4.20)}) describes the solution to equation for a particle which
moves in the magnetic field parallel to the coherent electromagnetic wave.
It is
the product of the oscillator eigenfunction for the ground state~\cite{12}
\begin{equation}
\varphi _{osc}=\pi ^{-1/4}\exp \left\{ -\frac{\left( \xi
-\sqrt{2<n>}
\right) ^2}2\right\}
\label{(4.21)}\end{equation}
and the solution

\begin{eqnarray}
\varphi _c&=&
\pi ^{-1/4}\left( 2^ss!\right) ^{-1/2}
\nonumber\\&&\hspace{0.1in}
\times\ H_s\left[ \sqrt{h}x_2-\frac{q_1}{\sqrt{h}}
+i\frac{a_1\sqrt{h}}{\sqrt{2}(h-Q)}(e+WHQ\frac{h^2+Q^2}{h^2-Q^2})\sin
(kx)\right] 
\nonumber\\&&\hspace{0.1in}
\times \exp
\left\{\rule{0mm}{4ex}\right.
i(px+q_1x_1)+iWH\frac{a_1}{\sqrt{2}}\frac{h^2x_2+q_1Q}{h+q}\sin (kx)
\nonumber\\&&\hspace{0.7in}
-WH\frac{h-Q}{e(h+Q)}\left( \frac{q_1}h-x_2\right)
\left(
h^2x_2+q_1Q\right)
\nonumber\\&&\hspace{0.7in}
+WH\frac{a_1}{\sqrt{2}}\frac{h^2x_2+q_1Q}{h+Q}\exp \left[
-i(kx)\right]
\left.\rule{0mm}{4ex}\right\},
 \label{(4.22)}
\end{eqnarray}
which corresponds
to an external magnetic field and a circularly
polarized classical
electromagnetic wave. Here we used that
$\xi \rightarrow
\sqrt{2<n>}$~\cite{11,12}.
The average energy density of the coherent
electromagnetic wave
is $k_0^2a_1^2/2=k_0<n>/V$. The solution of the equation in
$(2+1)$
dimensions for a composite scalar particle in the external coherent,
linearly polarized electromagnetic wave was constructed in~\cite{13}.
\Section{Conclusion}\label{sec:5}

We have obtained exact solutions to the
equation for a pion (or kaon or
another composite scalar particle) in the field
of a uniform static
magnetic field parallel to the direction of propagation of a
quantized
electromagnetic wave. As a particular cases we have a wave functions
for
point-like scalar particles (which are described by the Klein - Gordon
equation) in the field of a uniform static magnetic field and circularity
polarized $n$ photons. When the photon number goes to infinity with the
definite
density of photons, the wavefunction transforms to the solution
for the
particle in the magnetic field and coherent electromagnetic wave. In
the case of
very strong electromagnetic fields, the effects which are
connected with the
composite structure of particles can be sensed. As the
charge of the particles
$e$ enters the denominators of parameters~(\ref{(4.18)}) we
have nonperturbative
solutions which are impossible to receive using the
perturbative expansion in
$e$. Therefore, it is possible to use found
solutions for the investigation of
the nonperturbative effects in the strong
electromagnetic fields. The case of
the classical electromagnetic wave is
justified when numbers of photons in
strong light waves are large enough.
The exact solutions of the equation for a
composite scalar particle
interacting with the quantized fields are of some
interest. Such wavefunctions in quantized fields are useful by 
considering an interaction of a
particle with the field depending on time. In this case it is
impossible to
use the semiclassical scheme for determining absorbed or emitted
photons.
The found wavefunctions for particles in the quantized wave can be
applied
for this and for solving more complex problems.


\begin{thebibliography}{99}

\bibitem{1}  Kruglov, S. I. 1991 Sov. Phys.
J.\textbf{\ 34} 119; \textbf{34} 75;
1992 Sov. Phys. J. \textbf{35} 656; Lu J-F
, Kruglov S I 1994 J. Phys. G:
Nucl. Part. Phys. \textbf{20} 1017; Lu J-F,
Kruglov S I 1994 J. Math. Phys. 
\textbf{35} 4497.

\bibitem{2}  Kruglov S I
1995 J. Phys. G: Nucl. Part. Phys. \textbf{21} 1643.

\bibitem{3}  Kruglov S I
1996 J. Phys. G: Nucl. Part. Phys. \textbf{22} 461.

\bibitem{4}  Berson I Ya
1969 Izv. Acad. Nauk Latv. SSR, USSR, \textbf{5} 3;
1970 Izv. Acad. Nauk Latv.
SSR, USSR \textbf{3 }4.

\bibitem{5}  Fedorov M\ V, Kazakov A E 1973 Z. Physik
\textbf{261} 191.

\bibitem{6}  Abakarov D I, Oleynik V P 1972 The Journ. of
Theor. and Math.
Phys., USSR \textbf{12} 78.

\bibitem{7}  Petrun'kin V A 1981
Sov. Part. Nucl. \textbf{12} 278.

\bibitem{8}  L'vov A I 1993 Int. J. Mod.
Phys. \textbf{A8} 5267.

\bibitem{9}  Berson I YA 1969 Sov. J. JETP \textbf{56
}1627.

\bibitem{10}  Bateman H and Erdelyi A 1953 Higher Transcendental
Functions
(New York: McGraw-Hill).

\bibitem{11}  Glauber R J 1963 Rhys. Rev.
\textbf{130} 2529; \textbf{131} 2766.

\bibitem{12}  Schiff L I 1955 Quantum
Mechanics (New York: McGrow-Hill).

\bibitem{13}  Kruglov S I 1997 Mod. Phys.
Lett. \textbf{A12} 1699.

\end{thebibliography}
\end{document}